\documentclass[aps,twocolumn,showpacs]{revtex4}
\newcommand{\sect}[1]{\setcounter{equation}{0}\section{#1}}

\begin{document}
\title{Maximum entropy principle and power-law tailed distributions}
\author{G. Kaniadakis}
\email{giorgio.kaniadakis@polito.it} \affiliation{Dipartimento di
Fisica, Politecnico di Torino, \\ Corso Duca degli Abruzzi 24, 10129
Torino, Italy}
\date{\today}

\begin{abstract}
In ordinary statistical mechanics the Boltzmann-Shannon
entropy is related to the Maxwell-Bolzmann distribution
$p_i$ by means of a twofold link. The first link is differential and
is offered by the Jaynes Maximum Entropy Principle. Indeed, the
Maxwell-Boltzmann distribution is obtained by maximizing the Boltzmann-Shannon entropy under
proper constraints. The second link is algebraic and imposes that
both the entropy and the distribution must be expressed in
terms of the same function in direct and inverse form. Indeed, the
Maxwell-Boltzmann distribution $p_i$ is expressed in terms of the exponential
function, while the Boltzmann-Shannon entropy is defined as the mean value of
$-\ln(p_i)$.

In generalized statistical mechanics the second link is customarily
relaxed. Of course, the generalized exponential function defining
the probability distribution function after inversion, produces a
generalized logarithm $\Lambda(p_i)$. But, in general, the mean value
of $-\Lambda(p_i)$ is not the entropy of the system. Here we
reconsider the question first posed in [Phys. Rev. E {\bf 66},
056125 (2002) and {\bf 72}, 036108 (2005)], if and how is it
possible to select generalized statistical theories in which the
above mentioned twofold link between entropy and the distribution
function continues to hold, such as in the case of ordinary
statistical mechanics.

Within this scenario, apart from the standard
logarithmic-exponential functions that define ordinary statistical
mechanics, there emerge other new couples of direct-inverse
functions, i.e. generalized logarithms $\Lambda(x)$ and generalized
exponentials $\Lambda^{-1}(x)$, defining coherent and
self-consistent generalized statistical theories. Interestingly, all
these theories preserve the main features of ordinary statistical
mechanics, and predict distribution functions presenting power-law
tails. Furthermore, the obtained generalized entropies are both
thermodynamically and Lesche stable.

\end{abstract}
\pacs{05.90.+m, 05.20.-y, 51.10.+y, 03.30.+p}
\maketitle

\sect{Introduction}

Apart from the exponential distribution $\exp(-x)$, several distribution
functions have been considered both from the theoretical point of
view as well as to analyze experimental data. Ref. \cite{Mathai}
lists more than 130 different distributions functions, more or less
frequently used in statistical sciences.

In the last decades particular attention has been devoted to
distribution functions exhibiting power-law tails, namely
$A\,x^{-a}$ for $x \rightarrow \infty$, appearing in the
phenomenology of many physical, natural and artificial systems. The
power-law tailed distributions have been observed, for instance, in
high energy physics (plasmas \cite{Hasegawa}, cosmic rays
\cite{Vasyliunas,Biermann} and particle production processes
\cite{Wilk,Walton}), in condensed matter physics (anomalous diffusion
\cite{Ott,Bouchaud,Shlesinger}, fluid motion \cite{Solomon},
turbulence \cite{Antonia,Boghosian}), in natural sciences
(seismology \cite{Kasahara}, meteorology \cite{Ausloos} and
heliophysics \cite{Lu}), in biology (botany \cite{Niklas}, genomics
\cite{Nacher}), in economics (stock prices \cite{Gabaix, Plerou} and
personal annual income \cite{Pareto}), in sociodemography
(population of cities \cite{Blank}, frequencies of family names
\cite{Miyazima}, telephone calls \cite{Ebel}), in
linguistics \cite{Wichmann,Kosmidis}, in philology (classical mythology
\cite{Choi}), and also in fields involving human activity in general
(intensity of wars \cite{Roberts}, terrorism attacks \cite{Clauset},
citation of scientific papers \cite{Redner}, traffic in complex
networks \cite{Tadic}, etc.).

The very simple analytic expression for the tails i.e. $A\,x^{-a}$
in the above distributions, known as Pareto's or Zipf's law, has
been regularly adopted in the literature for data analysis
\cite{Newman,Sornette,Goldstein}. On the contrary, the question
regarding the form of the analytic expression for the power-law
tailed distribution that holds for $0<x<+\infty$ remains, at the
moment, an open problem.

In the last few years there have been studies of the general
properties of distribution functions of arbitrary forms
\cite{KLS04,KLS05,KL00,Abeg,Frank1}. On the other hand, the general
properties of an arbitrary entropic form are also well known
\cite{Csizar,Lesche,Abe2002}. Finally the connection bettween an
arbitrary entropy and the relevant distribution given through the
Jaynes maximum entropy principle \cite{Jaynes}, has been
considered by several authors \cite{KLS05,Abeg}.

A mechanism frequently used to explain the occurrence of statistical
distributions is based on evolution equations, essentially nonlinear
Fokker-Plank but also Boltzmann equations
\cite{BiroK,Polynomial-clas,Polynomial-quant,Fractional,Quons,
Curado,Chavanis,Frank2,PhA01,PLA01,PRE02,PRE05}. After fixing the
form of the evolution equation the entropy of the system is
automatically fixed and the probability distribution function is
obtained as a stationary state of the evolution equation.

Clearly the correctness of an analytic expression for a given power-law
tailed distribution, used to describe a
statistical system, is strongly related to the validity of the
generating mechanism.
In the present paper we will consider power-law tailed distribution
functions from a
general prospective and independently on the particular generating mechanisms.

We recall that, according to Maximum Entropy Principle, the Boltzmann-Shannon entropy $S=-\sum
\,p_{_i}\,\ln(p_{_i})$, yields the exponential Maxwell-Boltzmann distribution. In
generalized statistical mechanics, a special role is played by the
trace-form entropies
\begin{eqnarray}
S=-\langle\, \Lambda\, \rangle=-\sum_{i}\,p_{_i}\,\Lambda(p_{_i}) \
,\label{I1}
\end{eqnarray}
where $\Lambda(x)$ is an arbitrary, strictly increasing function
defined for $x>0$, which for $x \rightarrow 0^+$ becomes $-\infty$,
while $p=\{p_i\,, 1 \leq i \leq N\}$ is a discrete probability
distribution. The function $\Lambda(x)$ can be viewed as a generalization of the
ordinary logarithmic function. We remark that the entropy is the
ordinary mean value of $-\Lambda(p_i)$.  It is clear that the form
of $\Lambda(x)$ imposes the form of the distribution function
according to the maximum entropy principle.

In proposing generalized statistical theories some standard
properties (positivity, continuity, concavity, symmetry, etc.) are
customarily required for the entropy. Unfortunately, these
properties are not sufficiently strong to impose the form of the
entropy; then, the metaphor of ordinary statistical mechanics
remains the unique guiding principle for further generalization. In
this sense the maximum entropy principle, the cornerstone of
statistical physics, is a valid and powerful tool to explore new
roots in searching for generalized statistical theories.

The main goal of the present paper is to show that the maximum
entropy principle, by itself, suggests a generalization of ordinary
statistical mechanics. The obtained generalized statistical theories
preserve the main features of ordinary statistical mechanics and
predict distribution functions, showing power-law tails, some of
which are already known in the literature.

This paper is organized as follows:

In Sect. II we consider the maximum entropy principle within the
generalized statistical theories, based on trace form entropies.

In Sect. III we obtain a class of entropies depending on three real
parameters.

In Sect. IV  we consider the statistical distributions, related to
this class of three-parameter entropies, exhibiting power-law tails.

In Sect. V and Sect. VI we discuss the thermodynamic stability and
the Lesche stability of the generalized statistical theories, based
on the three-parameter entropies obtained in Sect. III.

In Sect. VII we show that the class of the three-parameter entropies
contains, as special cases, some already known in the literature, entropies.

In Sect. VIII we study a new, two-parameter subclass of entropies,
belonging to the three-parameter class of entropies.

In Sect. IX we make some concluding remarks.

Finally in the Appendix we solve a differential-functional equation
which permits us to obtain the three-parameter class of entropies.

\sect{The maximum entropy principle}

Let us consider the generalized entropy defined in Eq. (\ref{I1}).
We note that the particular probability distribution given by $p=\{\delta_{i a}\,, 1 \leq i \leq N \}$ where $a$ is a fixed integer with $1 \leq a \leq
N$, describes a state of the system for which we have the maximum
information. For this state we set $S=0$ and this condition imposes
for the generalized logarithm (i) $\Lambda(1)=0$ and (ii)
$0^+\,\Lambda(0^+)=0$. Furthermore, in analogy with the ordinary
logarithm we normalize the generalized logarithm through (iii) $
\Lambda'(1)\!=\!1$.

We introduce the constraints functional
\begin{eqnarray}
C={\rm a_{_0}}\!\!\left[\,\sum_{i}\, p_i-1 \right] + {\bf
a}\left[{\bf M} - \sum_{i}\, \, {\bf g}(i)\,p_i \right] \, ,
\label{II1}
\end{eqnarray}
where ${\rm a_{_0}}$  and ${\bf a}=\{{\rm
a_{_1},\,a_{_2},...,\,a_l}$\} are the $l+1$ Lagrange multipliers,
while ${\bf g}(i)=\{{\rm g_{_1}(i),\,g_{_2}(i),...,\,g_l(i)}$\}
are the generator functions of the $l$ moments ${\bf M}=\{{\rm
M}_1, {\rm M}_2,...,{\rm M}_l\}$.

The variational equation
\begin{eqnarray}
\frac{\delta }{\delta p_i}\,\, \big(\,S+C\,\big)=0 \ , \label{II2}
\end{eqnarray}
implies the maximization of the entropy $S$ under the constraints,
imposing the conservation of the norm of $p$
\begin{eqnarray}
\sum_i \, p_i=1  \ , \label{II3}
\end{eqnarray}
and the \textit{a priori} knowledge of the values  of the $l$
moments
\begin{eqnarray}
{\bf M}=\sum_i \,{\bf g}(i)\,p_i  \ . \label{II4}
\end{eqnarray}
Eq. (\ref{II2}) represents the maximum entropy principle and yields
the relationship
\begin{eqnarray}
\frac{\partial }{\partial p_i}\,p_i\,\Lambda(p_i)= - {\bf a \cdot
g }(i)+{\rm a}_{_0}  \ , \label{II5}
\end{eqnarray}
defining unambiguously the distribution function $p_i$.

We explain the meaning of the Lagrange multipliers ${\rm a}_j$ by
observing that ${\rm a}_j={\rm a}_j({\rm M}_1,\,...\,,\,{\rm M}_l)$
with $j=0,1,...,l$ and $S=S{_{{\rm M}_1,\,...\,,\,{\rm M}_l}}$
\cite{Frank1}. In the first step we multiply Eq. (\ref{II5}) by $p_i$,
then sum with respect to the index $i$, and finally we derive with
respect to ${\rm M}_k$ with $k=1,...,l$, obtaining
\begin{eqnarray}
&&\sum_i \,\frac{\partial p_i}{\partial {\rm M}_k}\,\frac{\partial
}{\partial p_i}\,p_i\,\Lambda(p_i)+ \sum_i \, p_i \,\frac{\partial
}{\partial {\rm M}_k} \, \frac{\partial }{\partial p_i}\,p_i\,\Lambda(p_i)\nonumber \\
&&- \frac{\partial {\rm a}_{0}}{\partial {\rm M}_k}+ \frac{\partial
}{\partial {\rm M}_k}\sum_{j=1}^{l}{\rm a}_j\,{\rm M}_j=0  \ ,
\label{II6}
\end{eqnarray}
and then
\begin{equation}
\frac{\partial S}{\partial {\rm M}_k}\!-\! \sum_i p_i \frac{\partial
}{\partial {\rm M}_k}  \frac{\partial}{\partial p_i}p_i\Lambda(p_i)
\!+\! \frac{\partial {\rm a}_{0}}{\partial {\rm M}_k}\!-\!
\frac{\partial }{\partial {\rm M}_k}\sum_{j=1}^{l}{\rm a}_j\,{\rm
M}_j\!=\!0 \ . \label{II7}
\end{equation}
In the second step we derive Eq. (\ref{II5}) with respect to ${\rm
M}_k$, subsequently multiply by $p_i$ and then sum with respect to the
index $i$, obtaining
\begin{eqnarray}
\sum_i p_i \,\frac{\partial }{\partial {\rm M}_k} \,
\frac{\partial}{\partial p_i}p_i\,\Lambda(p_i) \!-\! \frac{\partial
{\rm a}_{0}}{\partial {\rm M}_k}\!+\! \sum_{j=1}^{l}\frac{\partial
{\rm a}_j}{\partial {M}_k}{\rm M}_j\!=\!0 \ . \label{II8}
\end{eqnarray}
Finally, by combining Eqs. (\ref{II7}) and (\ref{II8}) we have
\begin{eqnarray}
\frac{\partial S}{\partial {\rm M}_k}&&=
-\sum_{j=1}^{l}\left(\frac{\partial }{\partial {\rm M}_k}\,{\rm
a}_j\,{\rm M}_j -\frac{\partial {\rm a}_j}{\partial {\rm M}_k}\,{\rm
M}_j\right)  \nonumber \\ &&=- \sum_{j=1}^{l}{\rm a}_j\,
\frac{\partial {\rm M}_j }{\partial {\rm M}_k}=- \sum_{j=1}^{l}{\rm
a}_j\, \delta_{jk}=-{\rm a}_k \ . \ \ \ \label{II10}
\end{eqnarray}
Then, for $j=1,...,l$ it holds
\begin{eqnarray}
{\rm a}_j=-\frac{\partial S}{\partial {\rm M}_j} \ , \label{II10}
\end{eqnarray}
regardless on the form of the generalized entropy.

\sect{Three-parameter entropies}

In the following, we are interested to the particular class of
probability distribution functions which are given in terms of the generalized exponential function
\begin{eqnarray}
{\cal E}(x)=\Lambda^{-1}(x) \ , \label{III2}
\end{eqnarray}
namely of the form
\begin{eqnarray}
p_i= \alpha \,{\cal E}\left(- \frac{{\bf a \cdot g }(i) -{\rm
a}_{_0}+\eta}{\lambda}\right) \ , \label{III4}
\end{eqnarray}
$\alpha, \lambda$ and $\eta$ being three scaling parameters.  We note that the parameters $\lambda$ and $\eta$ produce a scaling to the Lagrange multipliers.

Ordinary statistical mechanics corresponds to
the choice $\Lambda (x)=\ln(x)$, ${\cal E}(x)=\exp(x)$ and
$\{\alpha=1,\, \lambda=1, \, \eta=1\}$, or alternatively
$\{\alpha=1/e, \, \lambda=1, \, \eta=0\}$. We pose now the question regarding the possible existence
of further couples of functions $\Lambda (x)$, ${\cal E}(x)$ and set
of parameters $\{\alpha, \, \lambda, \, \eta\}$ defining statistical
theories, through Eqs. (\ref{I1}), (\ref{II5}) and (\ref{III4}), different from ordinary statistical mechanics.

Regarding this proposition we note that Eq.(\ref{III4}), after inversion, can be
written in the form
\begin{eqnarray}
\lambda \,\Lambda(p_i/\alpha)+\eta= - {\bf a \cdot g }(i) +{\rm
a}_{_0} \ , \label{III5}
\end{eqnarray}
which after comparison with Eq. (\ref{II5}) yields
\begin{eqnarray}
\frac{\partial }{\partial p_i} \,p_i \Lambda(p_i) = \lambda\,\,
\Lambda (p_i/\alpha)+ \eta \ . \label{III6}
\end{eqnarray}

The latter equation is a first order differential-functional
equation whose solutions determine unambiguously the form of the
generalized logarithm after taking into account the conditions (i)
$\Lambda(1)=0$, (ii) $0^+\,\Lambda(0^+)=0^+$ and (iii) $
\Lambda'(1)\!=\!1$.

After obtaining the generalized logarithm and by inversion the
generalized exponential, through Eqs.(\ref{I1}) and (\ref{III4}),
the entropy and the probability distribution function are
unambiguously fixed.

Regarding the meaning of Eq. (\ref{III6}) we note that the equation
guarantees a twofold link between entropy and the probability
distribution function: firstly we have a differential link offered
by the maximum entropy principle through Eq. (\ref{II5});  secondly
we have an algebraic link offered by Eq. (\ref{I1}) and Eq.
(\ref{III4}) where the same function in direct and inverse form,
i.e. $\Lambda(x)$ and ${\cal E}(x)$, defines the entropy and the
probability distribution function according to the standard rules of ordinary statistical mechanics. Eq. (\ref{III6}) has been proposed in ref. \cite{PRE05}, while the special case corresponding to $\eta=0$ has been considered in ref. \cite{PRE02}. The solutions of Eq. (\ref{III6}) for the special case $\eta=0$, have been obtained in Ref. \cite{PRE02,KLS05}.

Preliminarily we note that Eq. (\ref{III6}) admits the classical
solution corresponding to $\Lambda (x)=\ln(x)$ and $\{\alpha=1,\,
\lambda=1, \, \eta=1\}$, or alternatively $\{\alpha=1/e, \,
\lambda=1, \, \eta=0\}$. The general solution of the Eq. (\ref{III6}), derived in the
Appendix, is given in terms of a three-parameter function
$\Lambda(p_i)= \ln_{\kappa \tau \varsigma} (p_i)$ having the form

\begin{eqnarray}
\ln_{\kappa \tau \varsigma} (x)= \frac{\varsigma^{\kappa}\,
x^{\tau+\kappa} -\varsigma^{-\kappa}\, x^{\tau-\kappa}
-\varsigma^{\kappa} +\varsigma^{-\kappa}
}{(\kappa+\tau)\varsigma^{\kappa}+ (\kappa-\tau)\varsigma^{-\kappa}}
\ . \ \ \ \label{III7}
\end{eqnarray}
The three parameters $\{ \kappa, \tau, \varsigma \}$ are related to
the parameters $\{ \alpha, \lambda, \eta \}$ of Eq. (\ref{III6})
through
\begin{eqnarray}
&&\alpha=\left(\frac{1+\tau-\kappa}{1+\tau+\kappa}\right)^{\frac{1}{2\,\kappa}}
\ , \label{III8}
\\ &&
\lambda=\frac{\big(1+\tau-\kappa\big)^{\frac{\tau+\kappa}{2\,\kappa}}}
{\big(1+\tau+\kappa\big)^{\frac{\tau-\kappa}{2\,\kappa}}} \ ,
\label{III9}
 \\
&&\eta=(\lambda-1)\,\frac{\varsigma^{\kappa}-\varsigma^{-\kappa}}
{(\kappa+\tau)\varsigma^{\kappa}+ (\kappa-\tau)\varsigma^{-\kappa}}
\ . \label{III10}
\end{eqnarray}
The classical solution $\Lambda(p_i)=\ln(p_i)$, $\alpha=1/e$,
$\lambda=1$, $\eta=0$ is contained as a limiting case $(\tau=0, \,
\kappa\rightarrow 0)$ in the above three parameter class of
solutions.

Let us consider the dual function of $f(x)$ defined as follows:
\begin{eqnarray}
\widehat{f}(x)= -f(1/x) \ . \label{III10a}
\end{eqnarray}
It is easy to verify that the dual of a three-parameter logarithm again results to be a new three-parameter logarithm, i.e.
\begin{eqnarray}
\widehat{\ln}_{\kappa \tau \varsigma}(x)=\ln_{\kappa \tau'
\varsigma'}(x) \ , \label{III10b}
\end{eqnarray}
where $\tau'=-\tau$ and $\varsigma'=1/\varsigma$. In general a
three-parameter logarithm is not self-dual, but the class of the
three-parameter logarithm is self-dual.

In the present context a special role is played by the scaling
operation.  Let us consider the class of the functions ${\cal
F}=\{f(x), \,\, x>0 \}$, satisfying the condition $f(1)=0$ and
$f'(1)=1$. Starting from the function $f(x)\in {\cal F}$ we define
the corresponding scaled function $_{\sigma}f(x)\in {\cal F}$ as
follows:
\begin{eqnarray}
_{\sigma}f(x)= \frac{f(\sigma x)-f(\sigma)}
{\sigma \, f'(\sigma)} \ , \label{III11}
\end{eqnarray}
where $\sigma>0$ is the scaling parameter.

It is remarkable that two consecutive scaling operations are equivalent to a unique scaling operation, i.e.
\begin{eqnarray}
_{\sigma_2}\,[\,_{\sigma_1}f(x)\,]=\,\,_{\sigma_3}f(x)  \
,\label{III11a}
\end{eqnarray}
where $\sigma_3=\sigma_2 \sigma_1$.

The scaling operation has the following group properties:
\begin{eqnarray}
&&_{1}f(x)=f(x)   \ , \label{III11b} \\
&&_{1/\sigma}[\,_{\sigma}f(x)\,]=\,f(x) \ \ , \label{III11c}  \\
&&_{\sigma_1}\,[\,_{\sigma_2}f(x)\,]=\,\,_{\sigma_2}\,[\,_{\sigma_1}f(x)\,]   \ . \label{III11d}
\end{eqnarray}

It is straightforward to show that the three-parameter
logarithm $\ln_{\kappa \tau \varsigma}(x)$ has the interesting
property
\begin{eqnarray}
_{\sigma}\ln_{\kappa \tau \varsigma}(x)=\ln_{\,\kappa \tau
\varsigma'} (x) \ , \label{III12}
\end{eqnarray}
where $\varsigma'=\sigma \,\varsigma$. The scaled function
$_{\sigma}\ln_{\kappa \tau \varsigma}(x)$ of a three-parameter
logarithm again results to be a three-parameter logarithm with the
same first two parameters, while the third parameter, given by
$\varsigma'=\sigma \,\varsigma$, is the scaling parameter.  In
general a particular three-parameter logarithm is not self-scaling,
but the class of the three-parameter logarithms $\ln_{\kappa \tau
\varsigma}(x)$ is self-scaling.

The explicit form of the three-parameter class of entropies is given by
\begin{equation}
S=-\sum_{i}\,p{_i}\, \ln_{\,\kappa \tau \varsigma} (p{_i}) \ .
\label{III17}
\end{equation}

\sect{power-law tailed distributions}

It is immediately possible to verify that $\ln_{\,\kappa \tau \varsigma}(x) \in
C^{\infty}(R^+)$ presents the following asymptotic power-law
behaviour:
\begin{eqnarray}
&&\!\!\!\!\!\!\!\!\!\!\!\!\!\!\!\!\!\!\ln_{\,\kappa \tau \varsigma}
(x){\atop\stackrel{\textstyle\sim}{\scriptstyle x\rightarrow
{0^+}}}- \frac{A }{ x^{|\kappa|-\tau}}  \ , \label{IV1}
\\
&&\!\!\!\!\!\!\!\!\!\!\!\!\!\!\!\!\!\!\ln_{\,\kappa \tau \varsigma}
(x){\atop\stackrel{\textstyle\sim}{\scriptstyle x\rightarrow
{+\infty}}} A\,x^{|\kappa|+\tau} \ , \label{IV2}
\end{eqnarray}
where $A=\left[\left(|\kappa|+\tau\right)\,\varsigma^{|\kappa|}+
\left(|\kappa|-\tau\right)\, \varsigma^{-|\kappa|}\right]^{-1}$. For
$\tau>|\kappa|-1$ we have $\lim_{x\to 0^+} \,\,x\, \ln_{\,\kappa
\tau \varsigma}(x) =0$ so that it results
\begin{eqnarray}
&&\int\limits_0\limits^1 \ln_{\,\kappa \tau \varsigma}(x)\,dx
=-\frac{1}{(1+\tau)^2 -\kappa^2}\times \nonumber \\
&&\times
\left[1+(\tau^2-\kappa^2)\,\frac{\varsigma^{\kappa}-\varsigma^{-\kappa}}
{\left(\kappa+\tau\right)\,\varsigma^{\kappa}+\left(\kappa-\tau\right)\,
\varsigma^{-\kappa}} \right] \ \ . \ \ \ \ \ \ \ \ \ \label{IV3}
\end{eqnarray}
From Eqs. (\ref{III10a}) and (\ref{III10b}) it follows
\begin{equation}
\ln_{\,\kappa \tau \varsigma} (1/x)=-\ln_{\,\kappa \,\tau' \varsigma'}
(x) \ , \label{IV4}
\end{equation}
whith $\tau'=-\tau$ and $\varsigma'=1/\varsigma$.

After noting that when $-|\kappa|\leq \tau \leq|\kappa|$ it holds
$d\ln_{\,\kappa \tau \varsigma} (x)/dx>0$, we can define the
three-parameter exponential $\exp_{\,\kappa \tau \varsigma}(x)$ as
the inverse function of $\ln_{\,\kappa \tau \varsigma}(x)$.
The properties of $\exp_{\,\kappa \tau \varsigma} (x)$ readily
follow from those of $\ln_{\,\kappa \tau \varsigma}(x)$. For
instance, the asymptotic power-law behavior of $\exp_{\,\kappa \tau
\varsigma}(x)$ is given by
\begin{eqnarray}
\exp_{\,\kappa \tau
\varsigma}(x){\atop\stackrel{\textstyle\sim}{\scriptstyle
x\rightarrow {\pm\infty}}} (\pm A\,x)^{\,1/(\tau\pm|\kappa|)} \ ,
\label{IV8}
\end{eqnarray}
while Eq.(\ref{IV4}) implies
\begin{equation}
\exp_{\,\kappa \tau \varsigma}(x) \,\exp_{\,\kappa \tau'
\varsigma'}(-x)=1 \ . \label{IV9}
\end{equation}

The probability distribution function can been obtained after maximization of the three-parameter entropy (\ref{III17}), according to maximum entropy principle
\begin{eqnarray}
p_i=\alpha \,\exp_{\,\kappa \tau \varsigma} \left(-\,\frac{{\bf a
\cdot g }(i)-{\rm a}_{_0}+\eta}{\lambda}\right) \ . \label{IV6}
\end{eqnarray}

We remark that Eqs. (\ref{III7}), (\ref{III17}) and (\ref{IV6}) define
a very wide class of statistical theories, candidates to generalize
the ordinary statistical mechanics after passing a series of
validity tests, some of which will be considered in the next
sections.

\sect{Thermodynamic stability}

We denote with $q=\{q_i\}$ the optimal distribution which according
to maximum entropy principle, is defined by the algebraic equation
(\ref{II5}), i.e.
\begin{eqnarray}
\frac{\partial }{\partial q_i}\,q_i\,\ln_{\,\kappa \tau
\varsigma}(q_i)= - {\bf a \cdot g }(i) + {\rm a}_{_0} \ . \label{V1}
\end{eqnarray}

After introducing the entropy density for a general probability
distribution function $p_i$ according to
\begin{eqnarray}
\sigma(p_i)=-p_i\,\ln_{\,\kappa \tau
\varsigma}(p_i) \ , \label{V1a}
\end{eqnarray}
the constrained entropy  $\Phi=S+C$ can be written as
\begin{eqnarray}
{\Phi}(p)=\sum_i \left(\sigma(p_i)-\frac{\partial
\sigma(q_i)}{\partial q_i}\,p_i\right)+{\rm a}_o-{\bf a
\cdot}{\,\bf M} \ , \label{V2}
\end{eqnarray}
while its maximum value is given by
\begin{equation}
\Phi(q)=\sum_i\left(\sigma(q_i)-\frac{\partial
\sigma(q_i)}{\partial q_i}\,q_i\right) +{\rm a}_o-{\bf a
\cdot}{\,\bf M} \ . \label{V3}
\end{equation}
In order to calculate the difference of the above two functionals
\begin{equation}
{\Phi}(p)\!-\!{\Phi}(q)\!=\!\!\sum_i\!\left[\,
\sigma(p_i)\!-\!\sigma(q_i)\!-\!\frac{\partial
\sigma(q_i)}{\partial q_i}(p_i\!-\!q_i)\right]  \ , \label{V4}
\end{equation}
when $p_i\approx q_i$, we perform the Taylor expansion
\begin{equation}
\sigma(p_i)\approx \sigma(q_i)+ \frac{\partial
\sigma(q_i)}{\partial q_i}(p_i-q_i) + \frac{1}{2}
\frac{\partial^2 \sigma(q_i)}{\partial q_i^2}(p_i-q_i)^2 ,
\label{V5}
\end{equation}
and obtain
\begin{eqnarray}
{ \Phi}(p)-{\Phi}(q)=\sum_i \frac{1}{2} \frac{\partial^2
\sigma(q_i)}{\partial q_i^2}(p_i-q_i)^2  \ . \label{V6}
\end{eqnarray}
At this point we recall that $\Phi(q_i)$ represents the maximum
value of $\Phi(p_i)$ so that it holds
\begin{eqnarray}
 {\Phi}(p)-{\Phi}(q)\leq 0  \ . \label{V7}
\end{eqnarray}
The latter inequality and Eq. (\ref{V6}) imply
\begin{eqnarray}
 \frac{\partial^2 \sigma(x)}{\partial x^2}\leq 0 \ \ ,  \  \label{V8}
\end{eqnarray}
for any value of $x$. After taking into account the definition of
the density entropy, we get the following property for the
generalized logarithm
\begin{eqnarray}
 \frac{\partial^2 }{\partial x^2}\,\, x \,
 \ln_{\,\kappa \tau \varsigma}(x)\geq 0  \ . \label{V9}
\end{eqnarray}
The latter relationship can be verified easily by starting with Eq.
(\ref{III6}) and after taking into account that for $-|\kappa|\leq \tau \leq|\kappa|$ it results
$d\ln_{\,\kappa \tau \varsigma} (x)/dx>0$. Therefore we can conclude that the system described by the entropy
(\ref{III17}) is thermodynamically stable for $-|\kappa|\leq \tau \leq|\kappa|$.

\sect{Lesche stability}

The Lesche stability condition \cite{KLS05,Abe2002,Lesche}, imposes
that any physically meaningful entropy, depending on a probability
distribution function, should exhibit a small relative error
\begin{equation}
R=\left|\frac{S(p)-S(q)}{S_{max}}\right|  \ , \label{VI1}
\end{equation}
with regard to small changes in the probability distributions
\begin{equation}
D=||p-q|| \ . \label{VI2}
\end{equation}
Mathematically this implies that for any $\varepsilon>0$ there
exists $\delta>0$ such that $R\leq \varepsilon$ holds for all the
distribution functions satisfying $D\leq \delta$. It is known that
the Lesche stability condition holds for the Boltzmann-Shannon
entropy.

In the following, by adopting the procedure of ref. \cite{KLS05}, we
will show that the Lesche stability condition holds for the entire
class of three-parameter entropies (\ref{III17}).

By using the procedure of ref. \cite{KLS05}, we introduce the auxiliary function
\begin{eqnarray}
A(p,\,s) \equiv \sum_{i=1}^N\left[p_{_i}- \alpha\,\exp_{\kappa \tau
\varsigma}\left(-\frac{s}{\lambda}\right)\right]_+ \ , \label{VI3}
\end{eqnarray}
where we have used the definitions $[x]_+ \equiv \max(x,0)$. The
definition of  $A(p,\,s)$ implies that for any $s$ the following
relationship holds
\begin{eqnarray}
\Big|A(p,\,s)-A(q,\,s)\Big|\leq \sum_{i=1}^N|p_{_i}-q_{_i}| \equiv
|\!|p-q|\!|_1 \ , \ \ \ \ \ \ \ \ \  \label{VI4}
\end{eqnarray}
while for $s\geq-\lambda\,\ln_{\,\kappa \tau \varsigma}(1/N)$ it
results
\begin{eqnarray}
\Big|A(p,\,s)-A(q,\,s)\Big|<N\, \alpha\,\exp_{\,\kappa \tau
\varsigma}\left(-\frac{s}{\lambda}\right) \ . \label{VI5}
\end{eqnarray}

After taking into account the definition of $A(p,s)$ and Eq.
(\ref{III6}), it is easy to verify that the entropy (\ref{III17}) can be written in
form
\begin{eqnarray}
S(p) = \int\limits_{-1}\limits^{+ \infty}
\left[1-A(p,\,s)\right]ds\,\,\,-1 -\eta \ . \label{VI6}
\end{eqnarray}

From the latter relationship it follows that the absolute difference
of the entropies of the two different distributions $p=\{p_{_i}\}$
and $q=\{q_{_i}\}$ verifies the inequality
\begin{eqnarray}
\nonumber \Big|S(p)-S(q)\Big|&=&
\Bigg|\int\limits_{-1}\limits^{+\infty}
\left[A(p,\,s)-A(q,\,s)\right]\,ds\,\Bigg| \\ \nonumber
 &&\leq\int\limits_{-1}\limits^\ell\Big|
A(p,\,s)-A(q,\,s)\Big|\,ds
\\ && +\int\limits_\ell\limits^{+\infty}\Big|
A(p,\,s)-A(q,\,s)\Big|\,ds \ . \ \ \ \ \  \label{VI7}
\end{eqnarray}
Choosing $ -\lambda\,\ln_{\,\kappa \tau \varsigma}(1/N)  \leq \ell
<+\infty$, by using Eq (\ref{VI4}) in the first integral and Eq.
(\ref{VI5}) in the second integral of Eq. (\ref{VI7}), we obtain
\begin{eqnarray}
\Big|S(p)-S(q)\Big| &\leq&
\left(\ell+1\right) \,|\!|p-q|\!|_1 \nonumber \\
&+&N\,\alpha\int\limits_\ell\limits^{+\infty} \exp_{\,\kappa \tau
\varsigma}\left(-\frac{s}{\lambda}\right)\,ds \ . \ \ \ \
\label{VI8}
\end{eqnarray}
In particular the latter relationship holds for $\ell=\bar{\ell}$
being
\begin{eqnarray}
\bar{\ell} =-\lambda\,\ln_{\,\kappa \tau
\varsigma}\left(\frac{|\!|p-q|\!|_1}{\alpha\,N}\right) \ \ ,
\label{VI9}
\end{eqnarray}
the values of $\ell$ that minimizes the r.h.s. of the expression, as
long as $\bar{\ell}\geq-\lambda\,
 \ln_{\,\kappa \tau \varsigma}(1/N)$, which is true when
\begin{eqnarray}
\delta=|\!|p-q|\!|_1\leq\alpha \ , \label{VI10}
\end{eqnarray}
i.e. for sufficiently close distributions, according to the
metric. After some simple algebra one obtains the relative
difference of entropies
\begin{equation}
\bigg|\frac{S(p)-S(q)}{S_{\rm max}}\bigg|\leq \epsilon(\,\delta\,
,\,N) \ \ \ \  \ , \label{VI11}
\end{equation}
and
\begin{equation}
\epsilon(\,\delta\, ,\,N)=\frac{\delta}{\ln_{\,\kappa \tau
\varsigma}(1/N)}\,\left[\ln_{\,\kappa \tau
\varsigma}\left(\delta/N\right)\!-\!1\!+\!\frac{\eta}{\lambda}\right]
\ , \label{VI12}
\end{equation}
where $S_{\rm max}\equiv -\ln_{\,\kappa \tau \varsigma}(1/N)$. After
taking into account that $\lim_{x\to 0^+} x\,\ln_{\,\kappa \tau
\varsigma}(x) =0$, we find
\begin{eqnarray}
\lim_{\delta\to 0^+}\epsilon(\,\delta,N)=0^+ \ , \label{VI13}
\end{eqnarray}
such that if the two distributions are sufficiently close,
$\delta\rightarrow 0^+$, the corresponding difference
 $\epsilon$ of entropies (\ref{III17}) can be made small at will
and then we can conclude that the entropy (\ref{III17}) is Lesche
stable.

In the thermodynamic limit $N\to\infty$ we have
\begin{equation}
\lim_{{N\to\infty}}\epsilon(\,\delta,N)=\epsilon(\delta)=\delta^{\,\,1-|\kappa|+\tau}
\ , \  \label{VI14}
\end{equation}
and results that $\lim_{\delta\to 0^+} \epsilon(\,\delta)=0^+$. Thus
the entropy (\ref{III17}) is Lesche stable also in the thermodynamic
limit.

\sect{Special cases}

In the present section we consider a few special cases of the three
parameter generalized logarithm (\ref{III7}), some of which are
already known in the literature. The ordinary logarithm, which is
self-dual and self-scaling, can be obtained as the limiting case
when $\kappa \rightarrow 0$ and $\tau=0$, independently on the value
of the parameter $\varsigma$, such that we can write
$\ln(x)=\ln_{\,0\, 0 \,\varsigma} (x)$.

{\bf i)} {\it Self-dual logarithm}: the choice corresponding to
$\tau=0$, $\varsigma=1$ and $-1<\kappa < 1$, yields the generalized
logarithm $\ln_{\kappa} (x)=\ln_{\,0 \,0\, 1}(x)$, which together
with its inverse function assume the form
\begin{eqnarray}
&&\ln_{\kappa} (x)=
{\displaystyle\frac{x^\kappa-x^{-\kappa}}{2\,\kappa}}  \ ,
\label{VII1}
\\
&&\exp_{\kappa}(x)= \left(\sqrt{1+\kappa^2 x^{\,2}}+\kappa
x\right)^{1/\kappa} \ . \label{VII2}
\end{eqnarray}
The self-duality of the above functions is expressed through the
relationships
\begin{eqnarray}
&&\ln_{\kappa}(1/x)=-\ln_{\kappa}(x) \ , \label{VII2a}
\\
&&\exp_{\kappa}(x)\exp_{\kappa}(-x)=1 \ . \label{VII2b}
\end{eqnarray}

It is remarkable that the self-duality property, if imposed to the
three-parameter logarithm, i.e. $\ln_{\kappa \tau
\varsigma}(1/x)=-\ln_{\kappa \tau \varsigma}(1/x)$, automatically
fixes the parameters $\tau=0$ and $\varsigma=1$, so that
$\ln_{\kappa}(x)$ remains the only one self-dual generalized
logarithm.

The following integral representation for $\ln_{\kappa}(x)$ holds
\begin{eqnarray}
\ln_{\kappa}(x)=\frac{1}{2}\int_{1/x}^{\,x}\frac{dt}{t^{1+\kappa}} \
, \label{VII2c}
\end{eqnarray}
which in the limit of $\kappa \rightarrow 0$ reduces to a well known property of the
ordinary logarithm.

The properties of the functions $\ln_{\kappa}(x)$ and
$\exp_{\kappa}(x)$ have been study extensively in the literature due
to their relevance in relativistic statistical mechanics. The
deformation mechanism introduced by the parameter $\kappa$ is
induced directly by the Lorentz transformations of Einstein's
special relativity. Indeed $1/\kappa$ is the reciprocal of light
speed while $1/\kappa^2$ is the particle rest energy, in
dimensionless forms. On the other hand the self-duality relation
given by Eq. (\ref{VII2b}) leads to the relativistic dispersion
relation, obtaining in this way a direct link with the microscopic
dynamics of the system. An important property of $\exp_{\kappa}(x)$
is given by
\begin{equation}
\exp_{\kappa}(q_{_A}\oplus
q_{_B})=\exp_{\kappa}(q_{_A})\exp_{\kappa}(q_{_B}) \ \ ,
\end{equation}
$q_{_A}\oplus q_{_B}$ being the additivity law of the dimensionless
relativistic momenta, defined through
\begin{equation}
q_{_A}\oplus q_{_B}= q_{_A}\sqrt{1+\kappa^2 q_{_B}^2} +
q_{_B}\sqrt{1+\kappa^2 q_{_A}^2} \ \ .
\end{equation}
A further property of $\exp_{\kappa}(x)$ is given by
\begin{equation}
\frac{d}{d_{\kappa}\,q}\,\exp_{\kappa}(q)=\exp_{\kappa}(q) \ \ ,
\end{equation}
where the $\kappa$-derivative
\begin{equation}
\frac{d}{d_{\kappa}\,q}= \sqrt{1+\kappa^2 q^2}\,\,\frac{d}{d\,q}\ \
,
\end{equation}
emerge within the Lorentz-invariant differential calculus.

In order to better explain how the functions $\exp_{\kappa}(x)$ and
$\ln_{\kappa}(x)$ can define a statistical theory we consider
briefly the paradigm of statistical mechanics. We indicate with
$W_i$ the microscopic energy, with $\beta$ the reciprocal of the
temperature and with $\mu$ the chemical potential, so that, the
Lagrange multipliers are given by $a_0=\beta \mu$ and $a_1=\beta$.
The system entropy is given by
\begin{equation}
S=-\sum_{i}\,p{_i}\, \ln_{\,\kappa} (p{_i}) \ ,
\label{VII2d}
\end{equation}
while the maximum entropy principle yields the probability distribution function
\begin{eqnarray}
p_i=\alpha \,\exp_{\,\kappa} \left(-\,\beta\,\frac{W_i-\mu}{\lambda}\right) \ , \label{VII2e}
\end{eqnarray}
$\mu$ being the normalization constant, while the expressions of the
parameters $\lambda$ and $\alpha$ are given by Eqs (\ref{III8}) and
(\ref{III9}) respectively, after posing $\tau=0$.

Statistical mechanics based on Eqs. (\ref{VII2d}) and (\ref{VII2e})
has been introduced in \cite{PhA01,PLA01}, while in
\cite{PRE02,PRE05} its relativistic origin has been shown.

In the last few years various authors have considered the
foundations of this statistical theory, e.g.,  H-theorem and
molecular chaos hypothesis \cite{Silva06A,Silva06B},  thermodynamic
stability \cite{Wada1,Wada2},  Lesche stability
\cite{KSPA04,AKSJPA04,Naudts1,Naudts2},  Legendre structure of
ensued thermodynamics \cite{ScarfoneWada},  Gibbs theorem
\cite{Yamano}, geometrical aspects of the theory \cite{Pistone} etc. On the other hand specific applications to
physical systems have been considered, e.g., cosmic rays
\cite{PRE02}, relativistic \cite{GuoRelativistic} and classical
\cite{GuoClassic} plasmas in the presence of external
electromagnetic fields, relaxation in relativistic plasmas under
wave-particle interactions \cite{Lapenta,Lapenta2009}, kinetics of
interacting atoms and photons \cite{Rossani}, particle systems in
external conservative force fields \cite{Silva2008}, astrophysical
systems \cite{Carvalho,Carvalho2}, quark-gluon plasma formation \cite{Tewel},
relativistic quantum hadrodynamics \cite{Pereira}, etc. Other
applications concern nonlinear diffusion \cite{WadaScarfone2009},
dynamical systems at the edge of chaos
\cite{Corradu,Tonelli,Celikoglu}, fractal systems \cite{Olemskoi},
random matrix theory \cite{AbulMagd,AbulMagd2009}, error theory
\cite{WadaSuyari06}, game theory \cite{Topsoe}, information theory
\cite{WadaSuyari07}, etc. Also applications to economic systems have
been considered, e.g., to study personal income distribution
\cite{Clementi,Clementi2008,ClementiJSTAT}
and to model deterministic heterogeneity in tastes and product
differentiation \cite{Rajaon,Rajaon2008} etc.

{\bf ii)} {\it Self-scaling logarithm}: The case corresponding to
$\tau=-\kappa=\alpha/2$, is undoubtedly the more discussed in
the literature and produces the following generalized self-scaling
logarithm and exponential:
\begin{eqnarray}
&&\ln_{\alpha}(x)= \frac{1-x^{-\alpha}}{\alpha} \ , \label{VII3} \\
&&\exp_{\alpha}(x)=(1-\alpha \,x)^{-1/\alpha} \ . \label{VII4}
\end{eqnarray}
The self-scaling property is given by
\begin{eqnarray}
_{\sigma}\ln_{\alpha}(x)= \ln_{\alpha}(x) \ \ . \label{VII4a}
\end{eqnarray}

Both functions (\ref{VII3}) and (\ref{VII4}) have been considered
first by Euler in 1779 and explicitly has remarked that the function
$(x^{\omega}-1)/\omega$ reduces to the natural logarithm in the
$\omega \rightarrow 0$ limit \cite{Euler}. Successively, the
function (\ref{VII4}) was adopted in 1908 by Gosset to
construct the Student distribution \cite{Gosset}, which in 1968 was
introduced into plasma physics by Vasyliunas \cite{Vasyliunas}. In
mathematical statistics the function (\ref{VII4}) has been used to
construct the Burr-distribution \cite{Burr}. On the other hand the function  (\ref{VII3}) was adopted in 1967 by Harvda e Charvat \cite{Harvda}
to propose a generalized entropy in information theory. Finally both the
functions (\ref{VII3}) and (\ref{VII4}) were adopted by Tsallis in
1988, using the new parameter $q=\alpha+1$ to define the q-logarithm
$\ln_{q}(x)$ and q-exponential $\exp_{q}(x)$,
\begin{eqnarray}
&&\ln_{q}(x)= \frac{x^{1-q}-1}{1-q} \ , \label{VII3b} \\
&&\exp_{q}(x)=(1+(1-q) \,x)^{1/(1-q)} \ , \label{VII4b}
\end{eqnarray}
in order to develop nonextensive statistical mechanics
\cite{Tsallis88,Tsallis94}.

{\bf iii)} The case corresponding to the choice $\kappa=(a-1/a)/2$,
 $\,\,\tau=(a+1/a)/2-1$ and $\varsigma=1$ defines a generalized
logarithm
\begin{eqnarray}
\ln_{a}(x)= \frac{x^{a-1}-x^{1/a -1}}{a-1/a} \ , \label{VII5}
\end{eqnarray}
leading, by means of Eq. (\ref{III17}), to an entropy introduced in
the literature in 1997 by Abe \cite{Abeq} presenting the symmetry $a
\leftrightarrow 1/a$.

 {\bf iv)} Only in very few
cases, the generalized logarithm given by Eq. (\ref{III7}), can be
inverted to obtain analytically the corresponding generalized
exponential. For instance the choice $\kappa=3\gamma/2$,
$\tau=\gamma/2$, $\varsigma=1$, yields an invertible generalized
logarithm
\begin{equation}
\ln_{_{{{\scriptstyle \gamma}}}} (x)=
{\displaystyle\frac{x^{2\gamma}-x^{-\gamma}}{3\gamma}} \ , \
\label{VII6}
\end{equation}
corresponding to the generalized exponential
\begin{eqnarray}
\exp_{\gamma}(x)= \left[\left(\frac{1+y}{2}\right)^{1/3}
+\left(\frac{1-y}{2}\right)^{1/3}\right]^{1/\gamma}  , \ \ \
\label{VII6a}
\end{eqnarray}
with $y=\sqrt{1-4\gamma^3x^3}$, \cite{KLS05}.

{\bf v)} The three-parameter class of generalized logarithm
(\ref{III7}) contains a two-parameter subclass, already known in the
literature, defined by posing $\varsigma=1$
\begin{equation}
\ln_{\kappa \tau 1} (x)=x^{\tau}\, \ln_{\kappa} (x) \ . \label{VII7}
\end{equation}
The latter subclass of two-parameter-logarithms is not self-scaling.
From Eq. (\ref{III12}) with $\sigma=1/\varsigma$, one obtains the
following relationships:
\begin{eqnarray}
&&\ln_{\,\kappa \tau \varsigma}(x) =\,\,
_{\varsigma}\ln_{\, \kappa \tau 1}(x)   \ , \label{III14} \\
&&\ln_{\,\kappa \tau 1}(x) = \,_{1/\varsigma}\ln_{\,\kappa \tau
\varsigma}(x)  \ . \label{III15}
\end{eqnarray}

After introducing the new parameters $\alpha$ and $\beta$ through
$\kappa=(\alpha-\beta)/2$ and $\tau=(\alpha+\beta)/2$, the two-parameter logarithm $\ln_{ \kappa \tau 1}
(x)=\ln_{\alpha \beta} (x)$ can be written in the form
\begin{equation}
\ln_{_{{{\scriptstyle \alpha \beta}}}} (x)=
{\displaystyle\frac{x^\alpha-x^{\beta}}{\alpha-\beta}}  \ ,
\label{VII8}
\end{equation}
which is more familiar in the literature and has been considered
firstly by Euler \cite{Euler}. In general this two-parameter
logarithm cannot be inverted analytically, thus it is impossible to
define explicitly the corresponding generalized exponential. The
function (\ref{VII8}) has been used to construct the two-parameter
entropies introduced in the information theory in 1975 by Mittal
\cite{Mittal}, and by Sharma and Taneja \cite{Sharma}. The
generalized logarithm (\ref{VII8}) was reconsidered in statistical
mechanics in 1998 by Borges and Roditi \cite{Borges} and recently in
refs. \cite{KLS04,KLS05}. Clearly the $\kappa$-logarithm does not
emerge easily from (\ref{VII8}) and probably this is the reason why
it has not been discussed in the literature before 2001.

{\bf vi)} The three-parameter class of logarithm (\ref{III7})
contains a second, new two-parameter subclass of logarithms, namely
the scaled $\kappa$-logarithms, obtained by posing $\tau=0$.  The
corresponding class of generalized entropies and distribution
functions will be discussed in the next section.

The one-parameter and two-parameter generalized logarithms, already
known in the literature and discussed in the present section,
emerging as special cases of the three-parameter logarithm
(\ref{III7}), are reported in the table.

\begin{table*}
\caption{\label{tab:table1} Generalized logarithms already known in
the literature, obtained here as special cases of the
three-parameter logarithm. In the third column are reported the
values of the parametes $\kappa$, $\tau$, and $\varsigma$ to insert
in the general expression of the three-parameter logarithm, to
obtain as special cases the zero-, one-, and two-parameter
logarithms.}

\begin{ruledtabular}
\begin{tabular}{cccc}
\\
 & generalized logarithm expression  & parameter values &introduced in ref. \\ \\
 \hline \\
 three-parameter logarithm
 & $\ln_{\kappa \tau \varsigma} (x)= \frac{\varsigma^{\kappa}\,
x^{\tau+\kappa} -\varsigma^{-\kappa}\, x^{\tau-\kappa}
-\varsigma^{\kappa} +\varsigma^{-\kappa}
}{(\kappa+\tau)\varsigma^{\kappa}+
(\kappa-\tau)\varsigma^{-\kappa}}$
 &
 & \cite{PRE05} \\
 (most general case) & & & \\ \\
 \hline \hline \\
 1. \, zero-parameter logarithm
 & $\ln(x)$
 & $\tau=0$, \, $\kappa \rightarrow 0$
 & Napier \\ \\ \hline \\
 2. \, one-parameter logarithm
 & $\ln_{\kappa} (x)= \frac{
x^{\kappa} - x^{-\kappa} }{2\kappa}$
 & $\tau=0$, $\varsigma=1$
 & \cite{PhA01}  \\ \\ \hline \\
 3. \, one-parameter logarithm
 & $\ln_{q} (x)= \frac{
x^{1-q}-1 }{1-q}$
 & $\kappa=\tau=(1-q)/2$
 & \cite{Tsallis94}   \\  \\
 \hline \\
  4. \, one-parameter logarithm
 & $\ln_{a} (x)= \frac{
x^{a-1}-x^{1/a-1}}{a-1/a}$
 & $\varsigma=1$, \, $\kappa=(1-1/a)/2$
 & \cite{Abeq} \\
&& $\tau=(1+1/a)/2-1$
\\ \\ \hline \\
  5. \, two-parameter logarithm
 & $\ln_{\alpha \beta} (x)= \frac{
x^{\alpha} - x^{\beta} }{\alpha - \beta}$
 & $\varsigma=1$, \, $\kappa=(\alpha -\beta)/2$  & \cite{Mittal,Sharma}
 \\ && $\tau=(\alpha +\beta)/2$
 \\ \\ \hline \\
 6. \, two-parameter logarithm
 & $\ln_{\kappa \varsigma} (x)= \frac{\varsigma^{\kappa}\,
x^{\kappa} -\varsigma^{-\kappa}\, x^{-\kappa} -\varsigma^{\kappa}
+\varsigma^{-\kappa} }{\kappa\,(\varsigma^{\kappa}+
\varsigma^{-\kappa})}$
 & $\tau=0$
 & \cite{PRE05}
 \\ \\
\end{tabular}
\end{ruledtabular}
\end{table*}

\sect{Scaled $\kappa$-statistical mechanics}

We consider here the two-parameter generalized logarithms
$\ln_{\kappa \varsigma}(x)=\ln_{\,\kappa\, 0 \,\varsigma}(x)$,
obtained by posing $\tau=0$ in Eq. (\ref{III7}). It is easy to
verify that $\ln_{\kappa \varsigma}(x)$ is the scaled function of
$\ln_{\kappa}(x)$. The following relationships hold
\begin{eqnarray}
&&\ln_{\kappa \varsigma}(x) =\, _{\varsigma}\ln_{\kappa}
(x) \ \ , \label{IX1}  \\
&&\ln_{\kappa}(x) =
\, _{1/\varsigma}\!\ln_{\,\kappa \,\varsigma}(x) \ \ ,  \label{IX2} \\
&&\ln_{\kappa}(x) = \ln_{\kappa \,1}(x) \ \ , \label{IX3}
\end{eqnarray}
and results $\ln_{\,0\, \varsigma} (x)=\ln(x)$. The scaled
$\kappa$-logarithm assumes the form
\begin{equation}
\ln_{\kappa \varsigma}(x)= \frac{\varsigma^{\kappa}\, x^{\kappa}
-\varsigma^{-\kappa}\, x^{-\kappa}
-\varsigma^{\kappa}+\varsigma^{-\kappa} }{\kappa\varsigma^{\kappa}+
\kappa\varsigma^{-\kappa}} \ \ ,  \label{IX5}
\end{equation}
and can also be written as
\begin{eqnarray}
\ln_{\kappa \varsigma} (x)= \frac{1}{a}\,\ln_{\kappa} (\varsigma x)-\frac{b}{a}
\ , \label{IX4}
\end{eqnarray}
where $a=\sqrt{1+\kappa^2[\ln_{\kappa}(\varsigma)]^2}$ and
$b=\ln_{\kappa}(\varsigma)$. An important property of the scaled
$\kappa$-logarithm is that its inverse function, the scaled
$\kappa$-exponential, exists and assumes a very simple form for any
value of ${\kappa}$ and $\varsigma$:
\begin{equation}
\exp_{\kappa \varsigma}(x)= \frac{1}{\varsigma}\exp_{\kappa}\!
\left(a x + b \right) \ . \label{IX6}
\end{equation}

The following Taylor expansions hold:
\begin{equation}
\ln_{\kappa \varsigma}(1+x)
{\atop\stackrel{\textstyle\sim}{\scriptstyle x\rightarrow {0}}}x-
\left(\!1-\frac{b}{a}\,\kappa^2\right) \frac{x^2}{2} \ , \label{IX7}
\end{equation}
\begin{equation}
\exp_{\kappa \varsigma}(x)
{\atop\stackrel{\textstyle\sim}{\scriptstyle x\rightarrow
{0}}}1+x+\left(\!1-\frac{b}{a}\,\kappa^2\right)\frac{x^2}{2} \ ,
\label{IX8}
\end{equation}
while the power-law asymptotic behaviours of $\ln_{\kappa
\varsigma}(x)$ and $\exp_{\kappa \varsigma}(x)$ are given by
\begin{eqnarray}
&&\ln_{\kappa \varsigma}(x)
{\atop\stackrel{\textstyle\sim}{\scriptstyle x\rightarrow
{0^+}}}-\frac{x^{-|\kappa|}}{2a|\kappa|}\,\,
 \ , \ \ \ \ \ \ \ \label{IX9} \\
&&\ln_{\kappa \varsigma}(x)
{\atop\stackrel{\textstyle\sim}{\scriptstyle x\rightarrow
{+\infty}}}\,\frac{x^{\,|\kappa|}}{2a|\kappa|}\,\, \ , \label{IX10}
\end{eqnarray}
\begin{eqnarray}
\exp_{\kappa \varsigma}(x)
{\atop\stackrel{\textstyle\sim}{\scriptstyle x\rightarrow
{\pm\infty}}} \left |2a \kappa \right|^{\pm1/|\kappa|} \ \ .
\label{IX11}
\end{eqnarray}

The new class of two-parameter entropies is given by
\begin{eqnarray}
S=-\sum_i\,\, p_i\,\,\ln_{\kappa \varsigma} (p_i) \ . \label{IX12}
\end{eqnarray}
Consequently the maximum entropy principle produces a distribution
function having the form
\begin{eqnarray}
p_i=\alpha\,\exp_{\kappa \varsigma}\!\left(- \frac{{\bf a \cdot g }(i) -{\rm
a}_{_0}+\eta}{\lambda}\right) \ , \label{IX13}
\end{eqnarray}
with
\begin{eqnarray}
&&\alpha=\left(\frac{1-\kappa}{1+\kappa}\right)^{1/2\kappa}
\ , \label{IX13a} \\
&&\lambda=\sqrt{1-\kappa^2} \ , \label{IX13b}
\\
&&\eta=\frac{\sqrt{1-\kappa^2}-1}{\kappa}\,
\frac{\varsigma^{\kappa}-\varsigma^{-\kappa}}
{\varsigma^{\kappa}+\varsigma^{-\kappa}} \ . \label{IX13c}
\end{eqnarray}

The $\kappa$-logarithm results to be the semi-sum of two scaled
$\kappa$-logarithms with scaling parameters $\varsigma$ and
$1/\varsigma$ respectively, i.e.
\begin{eqnarray}
\ln_{\kappa}(x) =\frac{1}{2}\left [\, \ln_{\kappa \,
\varsigma}(x)+\ln_{\kappa \, 1\!/\!\varsigma}(x)\right ] \ .
\label{IX24}
\end{eqnarray}

A further connection between the $\kappa$-logarithm and the scaled
$\kappa$-logarithm is given by
\begin{eqnarray}
\ln_{\kappa \varsigma}(x)=\left[1-\frac{b}{a}\,\right]\,
\ln_{\kappa}(x) + \frac{b}{a} \, \ln_{\,\kappa \,\varsigma'}(x) \
 \ ,  \label{IX19}
\end{eqnarray}
with $\varsigma'=1/\alpha$. The following two expressions of
$\ln_{\,\kappa \,\varsigma'}(x)$ hold
\begin{eqnarray}
&&\ln_{\,\kappa\,\varsigma'}(x)= \lambda \ln_{\kappa}(x/\alpha)-1
 \ , \label{IX17} \\
&&\ln_{\,\kappa\,\varsigma'}(x)= \frac{d}{dx}
\left[\,x\,\ln_{\kappa}(x)\right] -1
 \ . \label{IX18}
\end{eqnarray}

It is remarkable that $\ln_{\kappa \varsigma}(x)$ in the limits of
$\varsigma \rightarrow 0$ and $\varsigma \rightarrow \infty$ reduces
to the self-scaling $q-$logarithm given by $\ln_{q}
(x)=(1-x^{1-q})/(q-1)$. After posing $q=1+|\kappa|$ we obtain
\begin{eqnarray}
&&\ln_{q}(x) = \ln_{\,\kappa \,0}(x)
  \ ,  \label{IX20} \\
&&\ln_{2-q}(x) = \ln_{\,\kappa \, \infty}(x)
   \ . \label{IX21}
\end{eqnarray}

On the other hand Eqs. (\ref{IX24}) and  (\ref{IX19}) in the
$\varsigma \rightarrow 0$ limit, reduce to
\begin{eqnarray}
\ln_{\kappa}(x) =\frac{1}{2}\left [ \, \,
\ln_{q}(x)+\ln_{\,2-q}(x)\,\right ] \ , \ \ \ \ \label{IX23}
\end{eqnarray}
\begin{eqnarray}
\ln_{q}(x) =\left(\!1+\frac{1}{|\kappa|}\right) \ln_{\kappa}(x) -
\frac{1}{|\kappa|}\, \ln_{\,\kappa \, \varsigma'}(x) \
 \ , \label{IX22}
\end{eqnarray}
with $q=1+|\kappa|$ and $\varsigma'=1/\alpha$.

\sect{Conclusions}

As conclusions, we recall briefly the main results obtained here.
The requirement that the generalized trace-form entropy
\begin{eqnarray}
S(p\,)=-\,\langle\, \Lambda\,
\rangle=-\sum_{i}\,p_{_i}\,\Lambda(p_{_i}) \ ,\label{IX1}
\end{eqnarray}
after maximization, according to maximum entropy principle, yields a
probability distribution function given by
\begin{eqnarray}
p_i= \alpha \,\Lambda^{-1}\left(- \frac{{\bf a \cdot g }(i) -{\rm
a}_{_0}+\eta}{\lambda}\right) \ , \label{IX2}
\end{eqnarray}
is sufficient to fix the form of $\Lambda (x)$ and therefore the
form of the entropy and of the probability distribution function.

The obtained class of generalized logarithms, $\Lambda
(x)=\ln_{\,\kappa \tau \varsigma}(x)$, depends on three free
parameters according to
\begin{eqnarray}
\ln_{\,\kappa \tau \varsigma}(x)= \frac{\varsigma^{\kappa}\,
x^{\tau+\kappa} -\varsigma^{-\kappa}\, x^{\tau-\kappa}
-\varsigma^{\kappa}+\varsigma^{-\kappa}
}{(\kappa+\tau)\varsigma^{\kappa}+ (\kappa-\tau)\varsigma^{-\kappa}}
\ . \label{IX3}
\end{eqnarray}

The relevant three-parameter entropy is both thermodynamically
stable and Lesche stable and contains as special cases all the
one-parameter and two-parameter trace form entropies appeared in the literature.

The ensuing three-parameter probability distribution functions
represent the minimal deformation of the Maxwell-Boltzmann
exponential distribution compatible with the maximum entropy
principle and exhibit power-law tails.

\sect{Appendix}

Here we solve the equation
\begin{eqnarray}
\frac{\partial }{\partial x} \,x \Lambda(x) = \lambda\,\, \Lambda
(x/\alpha)+ \eta  \ , \label{A1}
\end{eqnarray}
with the conditions $\Lambda(1)=0$,
$\Lambda'(1)=1$, $0^+\,\Lambda(0^+)=0$ and
$\Lambda(0^+)<0$.

After expressing the function $\Lambda(x)$ in terms of the auxiliary function $L(x)$ according to
\begin{eqnarray}
\Lambda (x)= \frac{{L}(\varsigma x)-{L}(\varsigma)} {\varsigma \,
{L'}(\varsigma)}  \ \ . \label{A2}
\end{eqnarray}
and after setting
\begin{eqnarray}
\eta=(\lambda-1)\,\frac{{L}(\varsigma)}{\varsigma\, {L'}(\varsigma)}
\ , \  \label{A3}
\end{eqnarray}
we obtain the following equation for the function $L(x)$
\begin{eqnarray}
\frac{\partial }{\partial x} \,x \, {L}(x) = \lambda
\,\,{L}\!\left(\frac{ x}{\alpha}\right) \ . \label{A4}
\end{eqnarray}
The conditions associated to the function $L(x)$ easily follow from those related to the function $\Lambda(x)$, i.e. $L(1)=0$, $L'(1)=1$, $0^+\,L(0^+)=0$ and $L(0^+)<0$.

The general solution of Eq. (\ref{A4}) has been obtained in Ref.
\cite{KLS05}. In the following we recall briefly the main steps of
the solution procedure.

First we perform the change of variable $L(x)= (1 /
x)\,h(\lambda\,\alpha\,\ln x) $ and $
x=\exp\left(t/(\lambda\,\alpha)\right)$ and reduce Eq. (\ref{A4}) to
a first order homogeneous differential-difference equation,
belonging to the class of delay equations
\begin{equation}
\frac{d\,h(t)}{d\,t}-h(t-t_{_0})=0  \ , \label{A5}
\end{equation}
where $ t_{0}=\lambda\,\alpha\,\ln\alpha$. The general solution
of this equation is given by
\begin{equation}
h(t)=\sum_{i=1}^n\sum_{j=0}^{m_i-1}a_{_{ij}}(s_{_1},\,\cdots,\,s_{_n})\,t^j\,
e^{s_{_i}\,t}\ \ , \label{A6} \
\end{equation}
where $n$ is the number of real solutions $s_{_i}$ (of
multiplicity $m_{_i}$) of the characteristic equation
$s_{_i}-e^{-t_{_0}\,s_{_i}}=0$. In general the integration constants
$a_{_{ij}}$ depend on the parameter $s_{i}$. The latter algebraic
equation does not admit real solutions for $t_{_0}<-1/e$. For
$t_{_0}\geq -1/e$ the number and multiplicity of the real solutions
of the characteristic equation depend on the value of $ t_{0}$ according to  \\
(a) for $-1/e<t_{_0}<0$, $n=2$ and $m_{_i}=1$, \\ (b) for
$t_{_0}=-1/e$,
$n=1$ and $m=2$, \\ (c) for $t_{_0}\geq0$, $n=1$ and $m=1$.\\

The solution of (\ref{A4}) for the case (a) can be written in the
form
\begin{equation}
L(x)=A_{_1}(\kappa_{_1},\,\kappa_{_2})\,
x^{\kappa_{_1}}+A_{_2}(\kappa_{_1},\,\kappa_{_2})\,
x^{\kappa_{_2}} \ , \label{A7} \
\end{equation}
where $\kappa_{_i}=\lambda\,\alpha\,s_{_i}-1$ and
$A_{_i}(\kappa_{_1},\,\kappa_{_2})$ the integration constants. The
boundary conditions $L(1)=0$ and $L'(1)=1$ imply $A_{_1}=-A_{_2}$
and $\kappa_{_1}\,A_{_1}+
\kappa_{_2}\,A_{_2}=(\kappa_{_1}-\kappa_{_2})\,A_{_1}=1$
respectively. After introducing the new parameters
$\kappa=(\kappa_{_1}-\kappa_{_2})/2$ and
$\tau=(\kappa_{_1}+\kappa_{_2})/2$, we can write the solution as
follows
\begin{equation}
L(x)=x^\tau\, {\displaystyle\frac{x^\kappa-x^{-\kappa}}{2\,\kappa}}
\ . \label{A8}
\end{equation}
The condition $L(0^+)<0$ implies $\tau\leq|\kappa|$. From the
characteristic equations we obtain the system $1\pm
\tau+\kappa=\lambda\,\alpha^{-\tau\mp\kappa}$,
which can be solved to determine the two constants $\alpha$ and
$\lambda$. The constant $\eta$ follows from Eqs. (\ref{A3}) and
(\ref{A8}). We obtain
\begin{eqnarray}
&&\alpha=\left(\frac{1+\tau-\kappa}{1+\tau+\kappa}\right)^{\frac{1}{2\kappa}}
\ , \label{A9} \\
&&\lambda=\frac{\big(1+\tau-\kappa\big)^{\frac{\tau+\kappa}{2\kappa}}}
{\big(1+\tau+\kappa\big)^{\frac{\tau-\kappa}{2\kappa}}} \ ,
\label{A10}
\\
&&\eta=(\lambda-1)\,\frac{\varsigma^{\kappa}-\varsigma^{-\kappa}}
{(\kappa+\tau)\varsigma^{\kappa}+ (\kappa-\tau)\varsigma^{-\kappa}}
\ , \label{A11}
\end{eqnarray}
with $\tau>|\kappa|-1$. This latter condition guarantees that
$0^+L(0^+)=0$

Case (b) corresponds to the limit of case (a) $\kappa_{_1}\to
\kappa_{_2}$ producing $L(x)=x^\tau\,\ln x$.  Case (c) gives the
trivial solution $L(x)=a\,x^b$ with $b=\lambda\,\alpha\,s-1$ which
can not be used to define a generalized logarithm.

The most general expression of $L(x)$ is given by Eq.(\ref{A8}).
Finally, after taking into account Eq. (\ref{A2}) we can write the
explicit form of the generalized logarithm $\Lambda (x)$ depending
on the three parameters $\{\kappa, \tau, \varsigma \}$:
\begin{eqnarray}
\Lambda (x)= \frac{\varsigma^{\kappa}\, x^{\tau+\kappa}
-\varsigma^{-\kappa}\, x^{\tau-\kappa}
-\varsigma^{\kappa}+\varsigma^{-\kappa}
}{(\kappa+\tau)\varsigma^{\kappa}+ (\kappa-\tau)\varsigma^{-\kappa}}
\ . \label{A12}
\end{eqnarray}
The conditions $\Lambda(0^+)<0$ and $0^+\Lambda(0^+)=0$ imply that
$\tau\leq|\kappa|$ and $\tau>|\kappa|\!-\!1$ respectively.

\end{document}